\documentclass[preprintnumbers, showpacs, amsmath]{revtex4}
\usepackage{graphicx}
\usepackage{dcolumn}
\usepackage{bm}
\begin{document}
\title{Surface-induced cubic anisotropy in nanomagnets}
\author{H. Kachkachi}
\email{kachkach@physique.uvsq.fr; http://hamid.kachkachi.free.fr}
\affiliation{Groupe d'Etude de la Mati\`ere Condens\'ee,
CNRS UMR8634 - Universit\'e de Versailles St. Quentin,
45 av. des Etats-Unis, 78035 Versailles, France}
\author{E. Bonet}
\email{bonet@grenoble.cnrs.fr}
\affiliation{Laboratoire Louis N\'eel - CNRS, BP166, 38042 Grenoble Cedex 9, France}
\begin{abstract}
We investigate the effect of surface anisotropy in a spherical many-spin magnetic
nanoparticle. By computing minor loops, 2D and 3D energyscape, and by investigating the behavior of the net magnetization, we show that in the case of a not too strong surface anisotropy the behavior of the many-spin particle may be modeled by that of a macro-spin with  an effective energy containing a uniaxial and cubic anisotropy terms. This holds for both the transverse and N\'eel's surface anisotropy models.
\end{abstract}
\maketitle
\section{\label{sec:intro}Introduction}
A magnetic nanoparticle exhibits many interesting and challenging novel properties such as exponentially slow relaxation at low temperature and superparamagnetic behavior above some temperature that depends on the particle's size and its underlying material. The magnetization of a superparamagnetic particle shuttles in a fast motion between the various anisotropy-energy minima. The stability of the magnetization against this thermally-activated reversal has become a crucial
issue in fundamental research as well as in technological applications.
Controlling this behavior, in view of room temperature applications,
requires a fair understanding of the magnetization dynamics at the
nanosecond time scale.

There are two competing approaches to the study of the static and dynamic
properties of a nanoparticle. i) \textit{One-spin particle} (OSP): a
macroscopic approach that models a nanoparticle as a single magnetic moment,
assuming coherent rotation of all atomic magnetic moments, and is exemplified
by the Stoner-Wohlfarth model for statics and N\'{e}el-Brown model for
dynamics \cite{wernsdorfer01acp, bonetetal99prl}. ii) \textit{Many-spin particle} (MSP):
this microscopic approach involves the atomic magnetic moment with
continuous degrees of freedom as its building block. It allows taking
account of the local environment inside the particle, including the
microscopic interactions and single-site anisotropy \cite{kacgar05springer}.
This approach becomes necessary when dealing with a very small nanoparticle
because the spin non-collinearities induced by strong boundary effects and
surface anisotropy invalidate the coherent-rotation assumption. However,
investigating the dynamics of an MSP is a real challenge. Indeed, within
this approach one is faced with complex many-body aspects with the inherent
difficulties related with analysing the energyscape (location of the minima,
maxima, and saddle points of the energy potential). This analysis is
unavoidable since it is a crucial step in the calculation of the relaxation
time and thereby in the study of the magnetization stability against
thermally-activated reversal. One may then address the question as to
whether there exist some cases in which the full-fledged theory that has
been developed for the OSP approach [see \cite{cofkalwal04worldsc} and
references therein] can still be used to describe an MSP. However, avoiding
somehow the spin non-collinearities induced by surface/interface anisotropy
means that some price has to be paid. In Ref.~\cite{garkac03prl} it was
shown that when the surface anisotropy is much smaller than the exchange
coupling, and in the absence of core anisotropy, the surface anisotropy
contribution to the particle's energy is of $4^{\mathrm{th}}$-order in the
net magnetization components and $2^{\mathrm{sd}}$-order in the surface anisotropy
constant. This means that the energy of an MSP with relatively weak surface anisotropy
can be modeled by that of an OSP whose effective energy contains an additional cubic-anisotropy potential.

It is the purpose of this work to extend the (numerical) result of Ref.~\onlinecite{garkac03prl} to include core anisotropy. This is achieved
by computing minor hysteresis loops, net magnetization components as functions of the field, and energyscape in 2 and 3 dimensions.
We find that because of surface anisotropy the minimum defined by the core uniaxial anisotropy splits into 4 minima, reminiscent of cubic anisotropy. We then show that the energyscape can be modeled by an effective energy of the net particle's magnetization containing a uniaxial- and cubic-anisotropy terms.
We also show that this result holds for two different models of surface anisotropy (transverse and N\'eel).
\section{Model and computing method}
We consider a spherical particle of $\mathcal{N}$ spins cut from a cube of
side $N$ (i.e., $N-1$ atomic spacings) with simple cubic lattice structure.
Due to the underlying (discrete) lattice structure, the particle thus obtained is not a sphere with smooth boundary because its outer shell presents apices, steps, and facets, resulting in many sites with different coordination numbers.

Our model Hamiltonian is the (classical) anisotropic Dirac-Heisenberg model~\cite{kacdim02prb}
\begin{equation}
\mathcal{H} = -\sum\limits_{\left\langle i,j\right\rangle }J_{ij}\mathbf{s}_{i}\cdot
\mathbf{s}_{j}-(g\mu _{B})\mathbf{H}\cdot \sum\limits_{i=1}^{\mathcal{N}}\mathbf{s}
_{i} + \mathcal{H}_{an},
\end{equation}
where $\mathbf{s}_{i}$ is the unit spin vector on site $i$, $\mathbf{H}$ the
uniform magnetic field, $\mathcal{N}$ the total number of spins (core and surface),
and $J_{ij}(=J>0)$ the nearest-neighbor ferromagnetic exchange coupling. $%
\mathcal{H}_{an}$ is the uniaxial single-site anisotropy energy
\begin{equation}  \label{uaa}
\mathcal{H}_{an}=-\sum\limits_{i}K_{i}(\mathbf{s}_{i}\cdot \mathbf{e}_{i})^{2},
\end{equation}
with easy axis $\mathbf{e}_{i}$ and constant $K_{i}>0$. If the spin at site $%
i$ is in the core, the anisotropy axis $\mathbf{e}_{i}$ is taken along the
reference $z$ axis and $K_{i}=K_{c}$. For surface spins, this axis is along
the radial (i.e., transverse to the cluster surface) direction and $K_{i}=K_{s}$. In this case, the model in (\ref{uaa}) for surface spins is called the Transverse surface
anisotropy (TSA) model.
We use the more general model of transverse direction given by the gradient [the vector perpendicular to the isotimic surface $\psi=\mathrm{constant}$ defining the shape of the particle, e.g. a sphere or an ellipsoid], because in the case of a spherical particle the transverse and radial directions coincide, whereas for another geometry such as an ellipsoid they do not.

A more physically appealing microscopic model of surface anisotropy was
introduced by N\'eel \cite{nee53cras} with
\begin{equation}\label{NSA}
\mathcal{H}_{an}^\text{N\'eel} = 
\dfrac{K_s}{2}\sum\limits_{i}\sum\limits_{j=1}^{z_i}({\bf s}_{i}\cdot{\bf e}_{ij})^{2},
\end{equation}
where $z_i$ is the coordination number of site $i$ and ${\bf e}_{ij}={\bf
r}_{ij}/r_{ij}$ is the unit vector connecting the site $i$ to its nearest
neighbors. This model is more realistic since the anisotropy at a given
site occurs only when the latter loses some of its neighbors, i.e. when it
is located on the boundary.
The model in (\ref{NSA}) will be referred to as the N\'eel surface anisotropy (NSA) model \cite{garkac03prl}.

Qualitatively, the NSA model is not quite different from the TSA model. 
For example, consider a site $i$ sitting on a $[100]$ facet, e.g. in the upper most plane normal to the $z$ axis. It has $4$ neighbors on that facet and one below it along the $z$ axis. 
From (\ref{NSA}), the corresponding energy reads
\begin{widetext}
\begin{subequations}
\begin{eqnarray*}
\mathcal{H}_i^\mathrm{NSA} &=& K_s\left[ \left( \mathbf{s}_{i}\cdot \mathbf{e
}_{x}\right) ^{2} + \left( \mathbf{s}_{i}\cdot -\mathbf{e}_{x}\right)
^{2} + \left( \mathbf{s}_{i}\cdot \mathbf{e}_{y}\right)^{2}+\left( \mathbf{s}
_{i}\cdot-\mathbf{e}_{y}\right)^{2}
+ \left( \mathbf{s}_{i}\cdot -\mathbf{e}_{z}\right)^{2}\right] \\
&=& K_s\left[2s_{i,x}^{2} + 2s_{i,y}^{2} + s_{i,z}^{2}\right] 
= 2 K_s - K_s\,s_{i,z}^{2},
\end{eqnarray*}
\end{subequations}
\end{widetext}
where we have used $\parallel{\bf s}_i\parallel=1$. This implies that if $K_s>0$ the easy direction is along $\pm \mathbf{e}_{z}$, i.e., normal to the facet, and if $K_s<0$ the facet becomes an easy plane. 
Therefore, upon dropping the irrelevant constant, we rewrite the above energy as
\[
\mathcal{H}_i^\mathrm{NSA} = -K_s\,\left(\mathbf{s}_{i}\cdot \mathbf{e}_{z}\right)^{2}. 
\]
which is the same as the TSA in Eq.~(\ref{uaa}) for the site considered. 
More generally, averaging the NSA over a surface perpendicular to the direction $\mathbf{n}$ leads to [see Eqs.~(6, 7) in Ref.~\onlinecite{garkac03prl}]
\begin{equation}
\mathcal{H}_i^\mathrm{NSA} = - K_s\left(|n_x|s_x^2 + |n_y|s_y^2 + |n_z|s_z^2\right)
\end{equation}
thus favoring among $(x, y, z)$ the direction closest to the surface normal. This explains the similarity between the results obtained with TSA and NSA.
Using the components $|n_\alpha|,\alpha=x,y,z$, the atomic surface density reads \cite{garkac03prl}
\begin{equation}\label{fnDef}
f(\mathbf{n})=\max\left\{|n_{x}|,|n_{y}|,|n_{z}|\right\},  
\end{equation}
and thereby we have the particular cases
\begin{equation*}
\mathcal{H}_i^\mathrm{NSA} = 
\left\{
\begin{array}{ll}
-K_s\,s_{z}^{2}, & n_{z}=1 \\
-K_s(s_{x}^{2} + s_{y}^{2})/\sqrt{2}, & n_{x} = n_{y} = \frac{1}{\sqrt{2}} \\
-K_s/\sqrt{3}, & n_{x} = n_{y} = n_{z} = \frac{1}{\sqrt{3}}.
\end{array}
\right.
\end{equation*}
For comparison, taking account of the atomic surface density, the TSA model is described by
\begin{equation}\label{ESPerp}
\mathcal{H}_i^\mathrm{TSA} = - K_s\,(\mathbf{n}\cdot \mathbf{s}_i)^{2}\,f(\mathbf{n})  
\end{equation}
and the whole effect comes from the atomic surface density. 
Quantitatively, in the NSA model the effect is bigger, since for the surface cut perpendicular to the grand diagonal of the cubic lattice the anisotropy completely disappears, whereas in (\ref{ESPerp}) the surface anisotropy is only reduced by the factor $1/\sqrt{3}$.

Because we are dealing with an MSP, the energyscape cannot be represented in
terms of the coordinates of all spins. Instead, we may represent it in terms
of the coordinates of the particle's net magnetization. For this purpose, we
fix the global or net magnetization, $\mathbf{m}$, of the particle in a
desired direction ${\bm m}_{0}$ ($|{\bm m}_{0}|=1$) by using the energy
function with a Lagrange multiplier ${\bm \lambda }$ \cite{garkac03prl}:
\begin{equation}  \label{FFuncDef}
\mathcal{F}=\mathcal{H}-\mathcal{N}{\bm \lambda \cdot }\left( {\bm m}-{\bm m}%
_{0}\right) ,\qquad {\bm m\equiv }\frac{\sum_{i}\mathbf{s}_{i}}{\left|
\sum_{i}\mathbf{s}_{i}\right| }.
\end{equation}
To minimize $\mathcal{F},$ we solve the evolution equations
\begin{eqnarray}
\mathbf{\dot{s}}_{i} &=&-\left[ \mathbf{s}_{i}\times \left[ \mathbf{s}%
_{i}\times \mathbf{F}_{i}\right] \right] ,\qquad \mathbf{F}_{i}\equiv
-\partial \mathcal{F}/\partial \mathbf{s}_{i}  \nonumber  \label{LLEqs} \\
{\dot{\bm \lambda }} &=&\mathbf{\partial }\mathcal{F}/\partial {\bm \lambda
=-}\mathcal{N}\left( {\bm m}-{\bm m}_{0}\right) ,
\end{eqnarray}
starting from $\mathbf{s}_{i}={\bm m}_{0}=\mathbf{m}$ and ${\bm\lambda =0,}$
until a stationary state is reached. In this state ${\bm m}={\bm m}_{0}$ and
$\left[ \mathbf{s}_{i}\times \mathbf{F}_{i}\right] =0,$ i.e., the torque due
to the term $\mathcal{N}{\bm\lambda \cdot }\left( {\bm m}-{\bm m}_{0}\right)
$ in $\mathcal{F}$ compensates for the torque acting to rotate the global
magnetization towards the minimum-energy directions [see discussion in Ref.~\cite{garkac03prl}]. The orientation of the net magnetization is then given
either in Cartesian coordinates $(m_{x},m_{y},m_{z})$ or in spherical
coordinates $(\theta_{n},\varphi_{n})$.
\section{\label{sec:resultsdiscussion}Results and discussion}
The anisotropy is taken uniaxial in the core and along the $z$ axis. On the surface, we use the TSA model, except in Fig.~\ref{NSA360kc001ks03_p0p45} where the NSA model is used instead. All anisotropy constants are measured with respect to the exchange coupling $J$, so we define the reduced constants, $k_c\equiv K_C/J, k_s\equiv K_S/J$. In all calculations, unless otherwise specified, the magnetic field is applied at an angle of $\pi/4$ with respect to the core easy axis, i.e., $\mathbf{e}_z$ [see discussion at the end of section \ref{EOSP}].

To illustrate the issue stated in the abstract and at the end of section \ref{sec:intro}, we have chosen $k_c = 0.01, k_s = 0.3$, which in real units correspond to $K_c\simeq 1.3\times 10^{-23}$ J/atom and $K_s\simeq 4\times 10^{-22}$ J/atom. These values do not really correspond to some particular nanoparticle material, since for cobalt nanoparticles, for instance, one would have $k_c \simeq 0.0024$. However, they are consistent with the fact that i) $k_c$ is typically at least two orders of magnitude smaller than exchange, as has been estimated by many authors, ii) the surface anisotropy constant is about one order of magnitude larger than the core constant. Here we start with a slightly larger constant, but then it is varied [see Fig.~\ref{enscape2D_h0_varyks} below].

\begin{figure}[floatfix]
\begin{center}
\includegraphics[width=8cm]{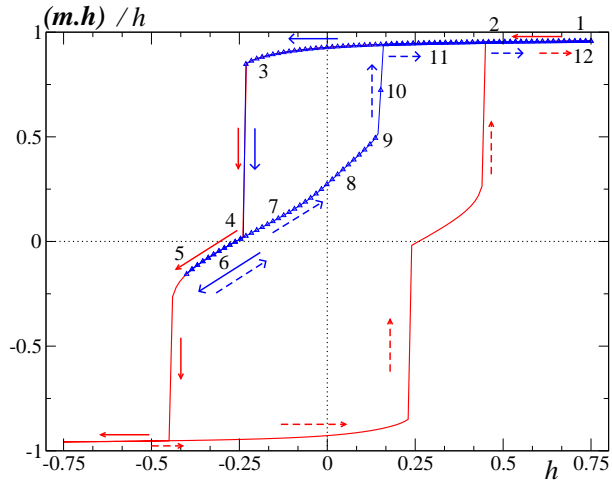}
\end{center}
\caption{(Color online) Full and minor hysteresis loops of a spherical particle of $%
\mathcal{N}=360$ spins with $k_c=0.01, k_s=0.3$. The return field is $h=-0.4$
($h=H/H_k,\,H_k=2K_c$).}
\label{loops_full_vs_minor}
\end{figure}

\begin{figure}[floatfix]
\begin{center}
\includegraphics[width=8cm]{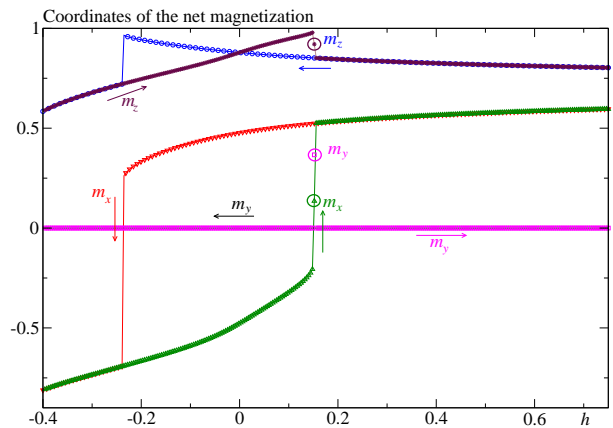}
\end{center}
\caption{(Color online) Cartesian coordinates of the net magnetization $\mathbf{m}$
computed along the minor hysteresis loop of Fig.~\ref{loops_full_vs_minor}.
The rings mark the unique state at which the magnetization goes out of the $%
x-z$ plane [see text]. }
\label{mag_cc}
\end{figure}

In Fig.~\ref{loops_full_vs_minor} we plot (in solid line) the full hysteresis cycle indicated by narrow-headed arrows, solid for the upper and dashed for the lower branch. In triangles is the cycle obtained upon decreasing the field from the positive saturation value down to $-0.4$ and then backwards.
%
%
It is clear that the appearance of minor loops is an indication of the
existence of several local minima induced by surface anisotropy. %
\begin{figure*}[floatfix]
\begin{center}
\includegraphics[angle=-90,width=2.5cm]{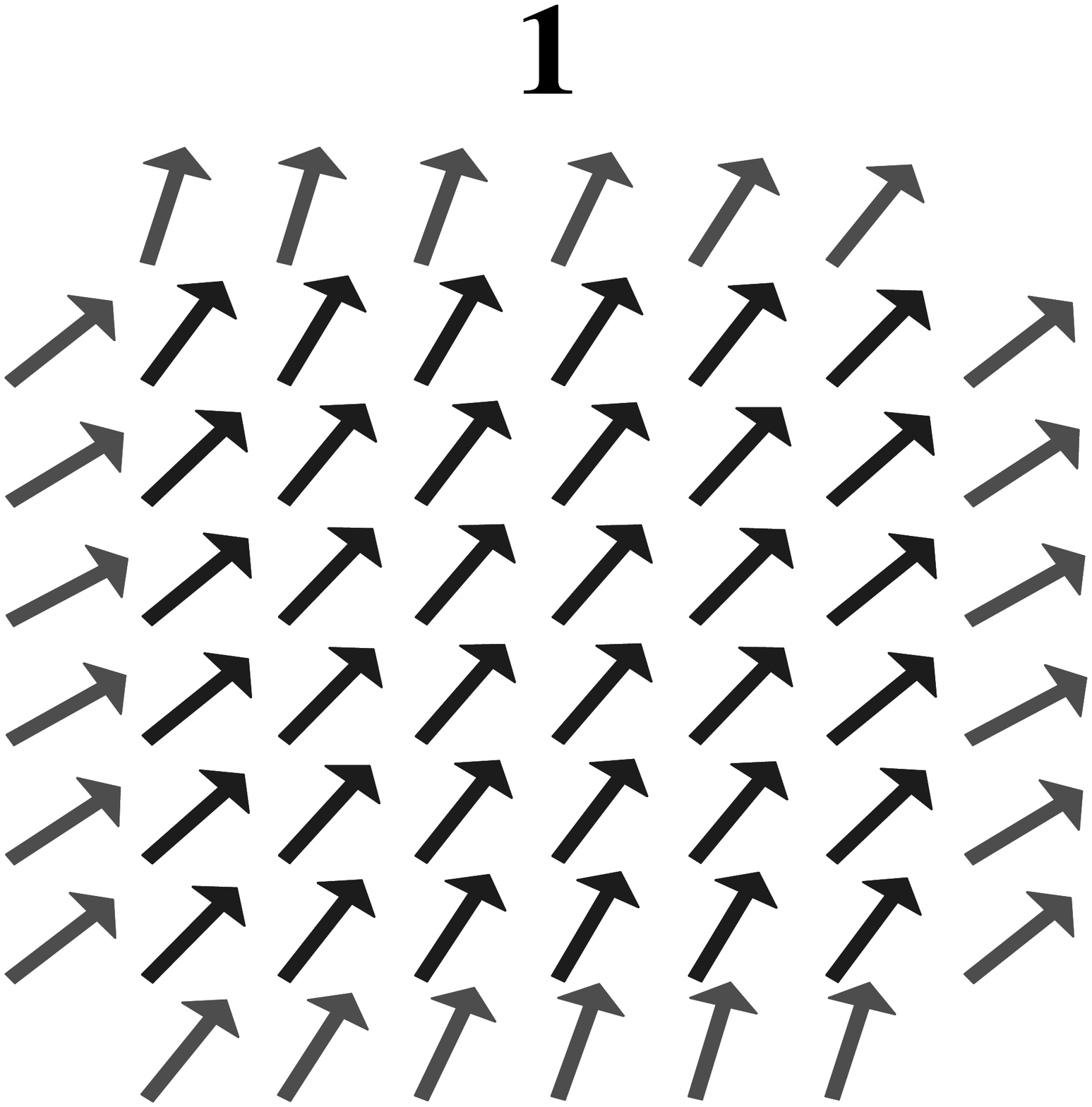}\hspace{%
0.25cm} \includegraphics[angle=-90,width=2.5cm]{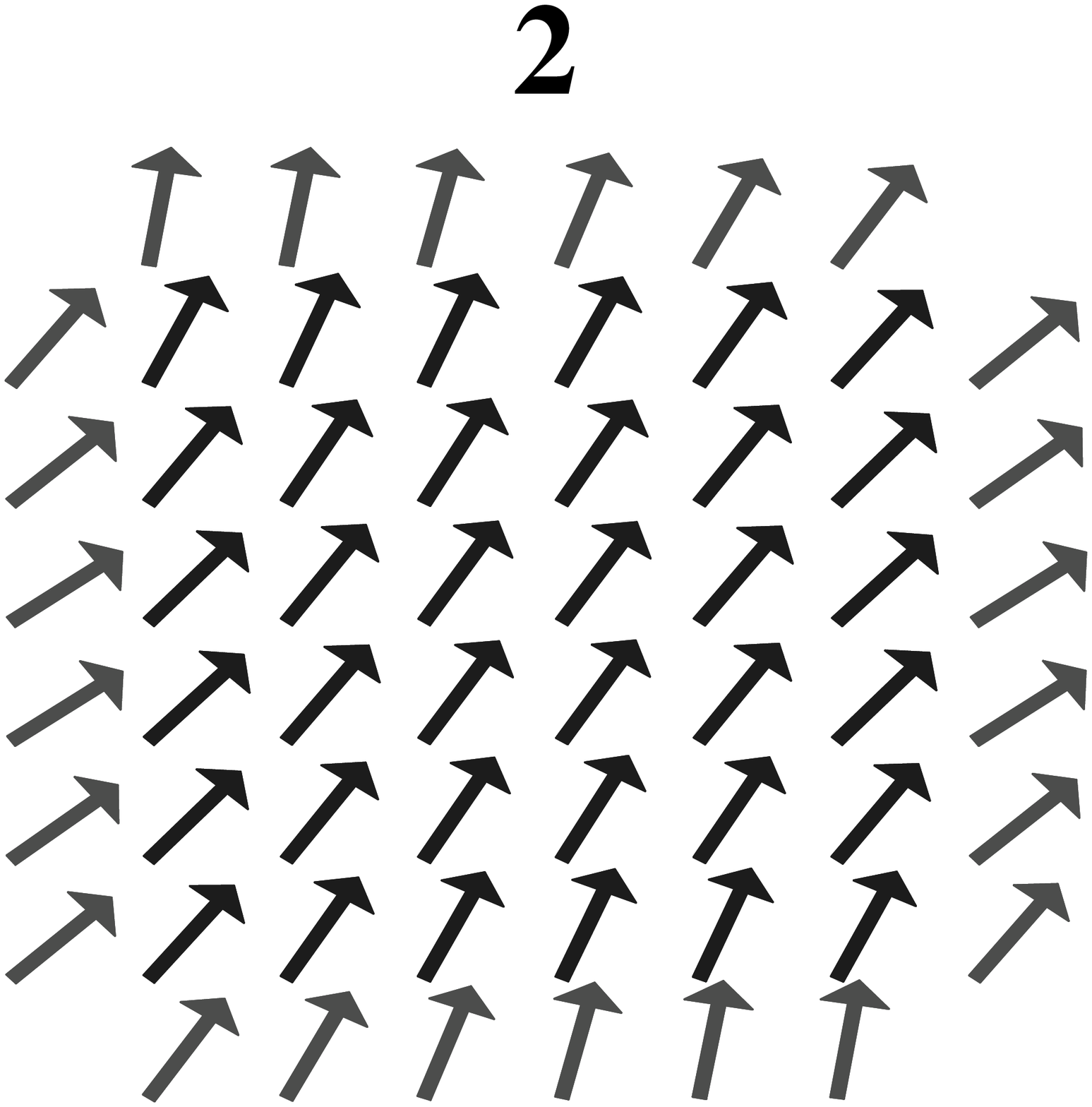}%
\hspace{0.25cm}
\includegraphics[angle=-90,width=2.5cm]{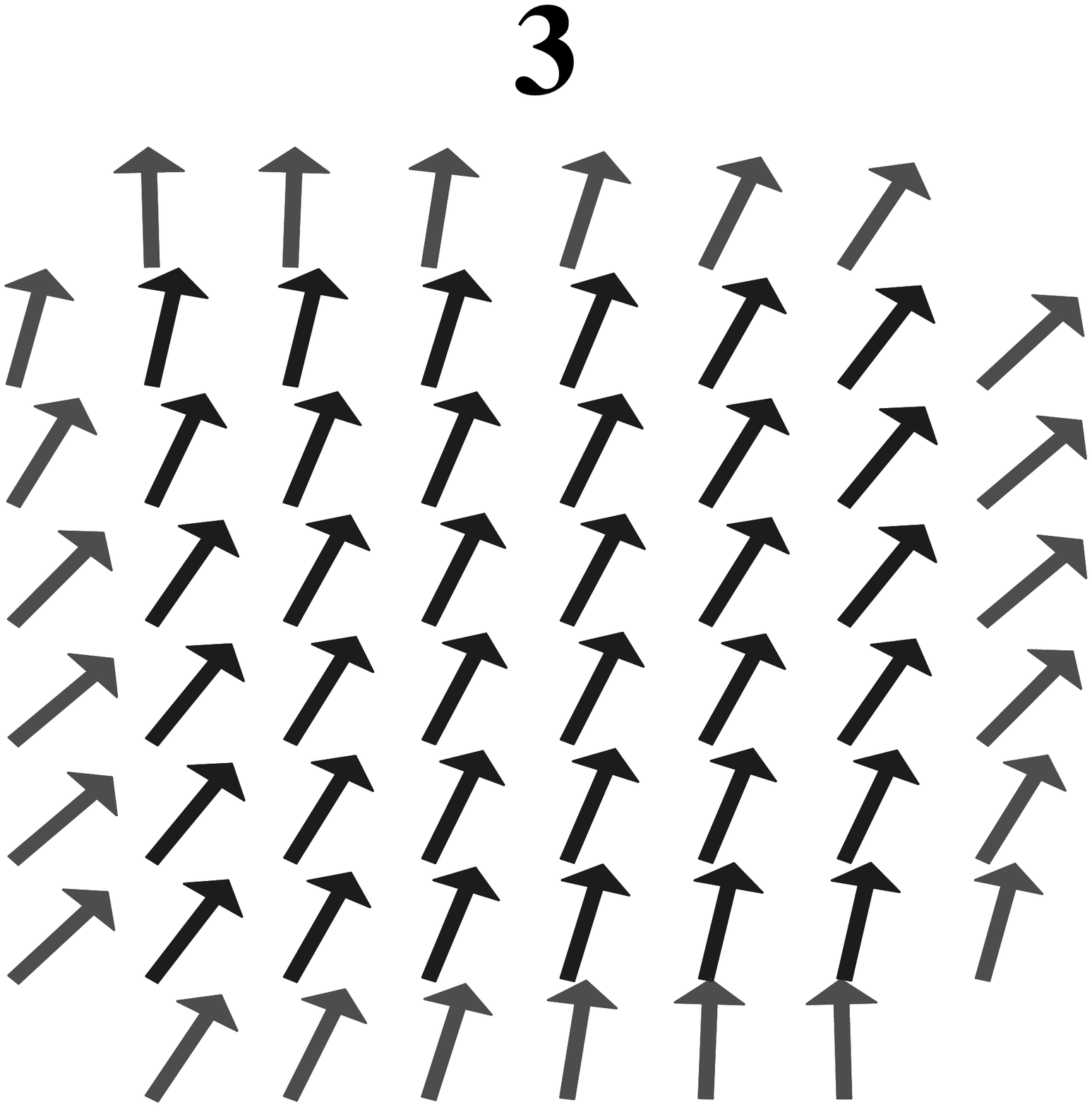}%
\hspace{0.25cm}
\includegraphics[angle=-90,width=2.5cm]{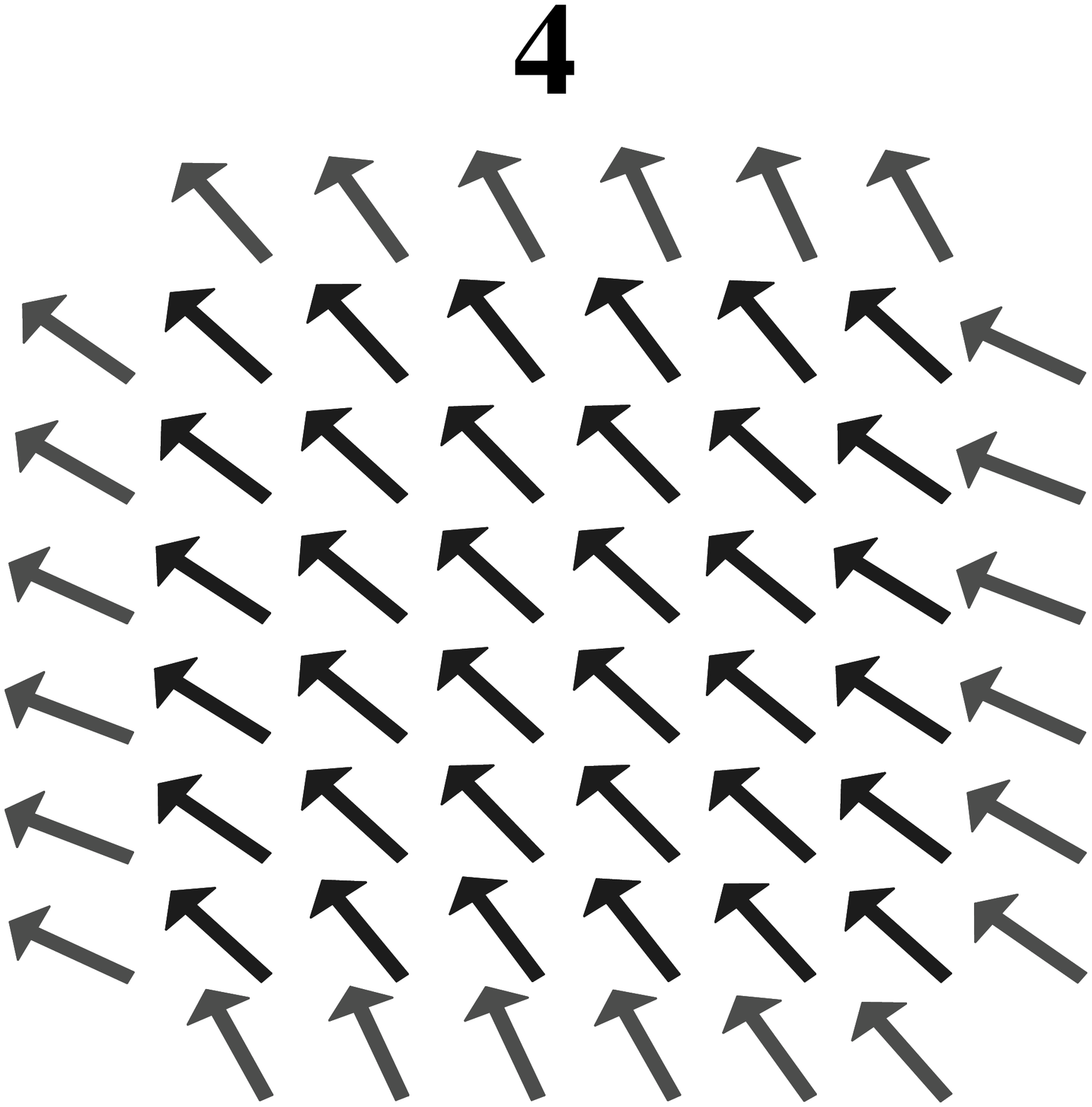}\\[%
0pt]
\vspace{0.5cm}
\includegraphics[angle=-90,width=2.5cm]{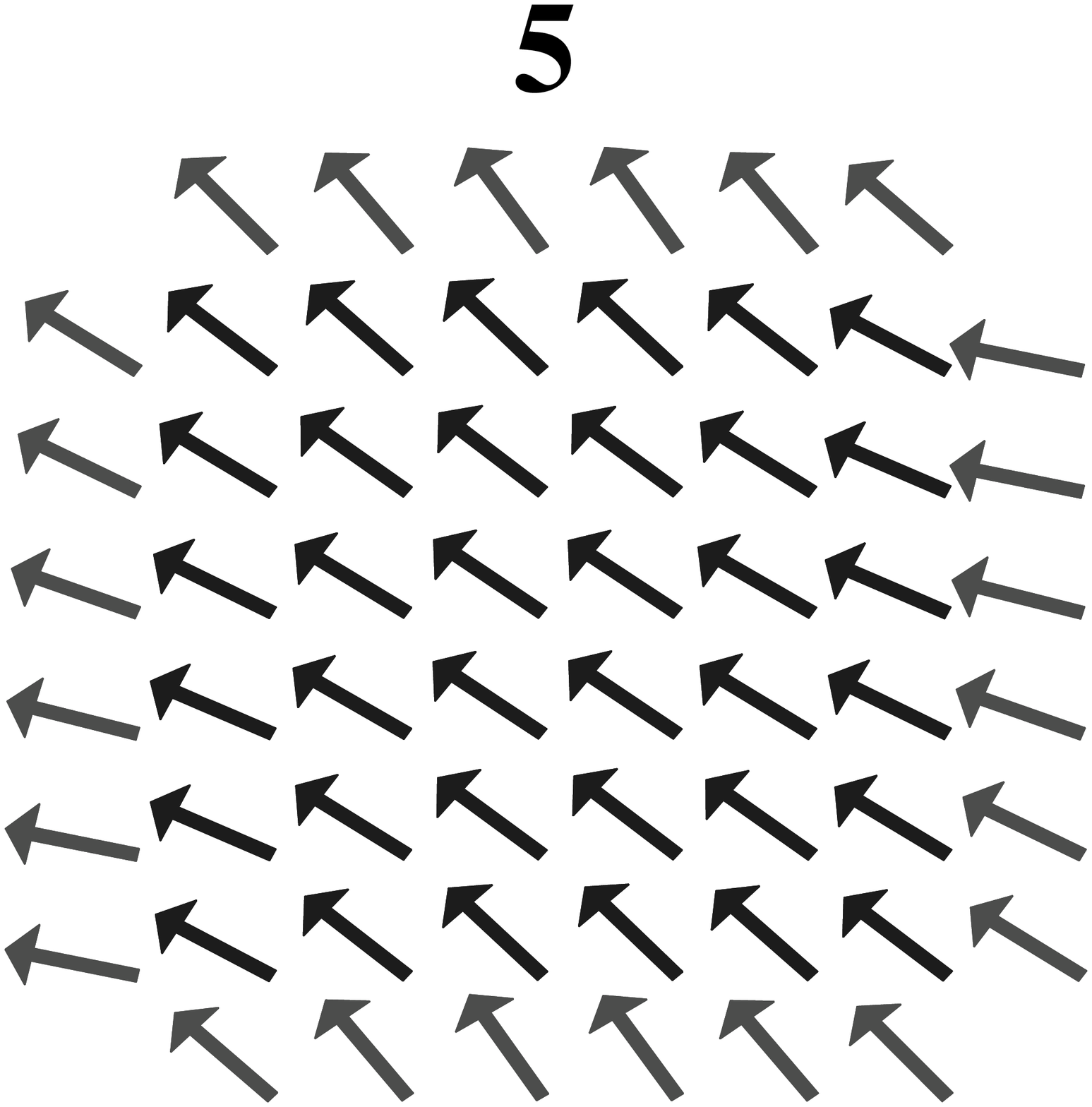}%
\hspace{0.25cm}
\includegraphics[angle=-90,width=2.5cm]{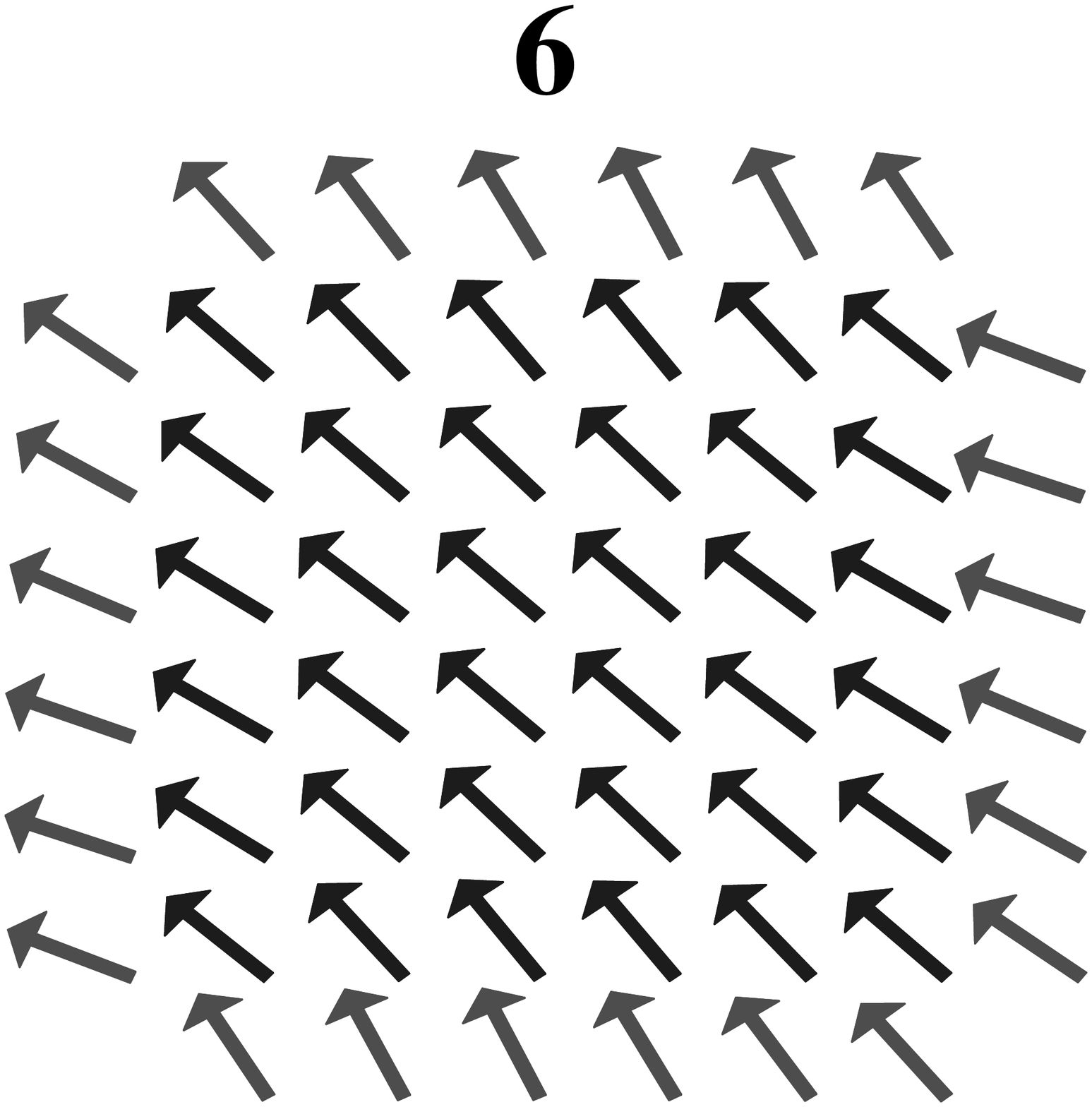}\hspace{%
0.25cm} \includegraphics[angle=-90,width=2.5cm]{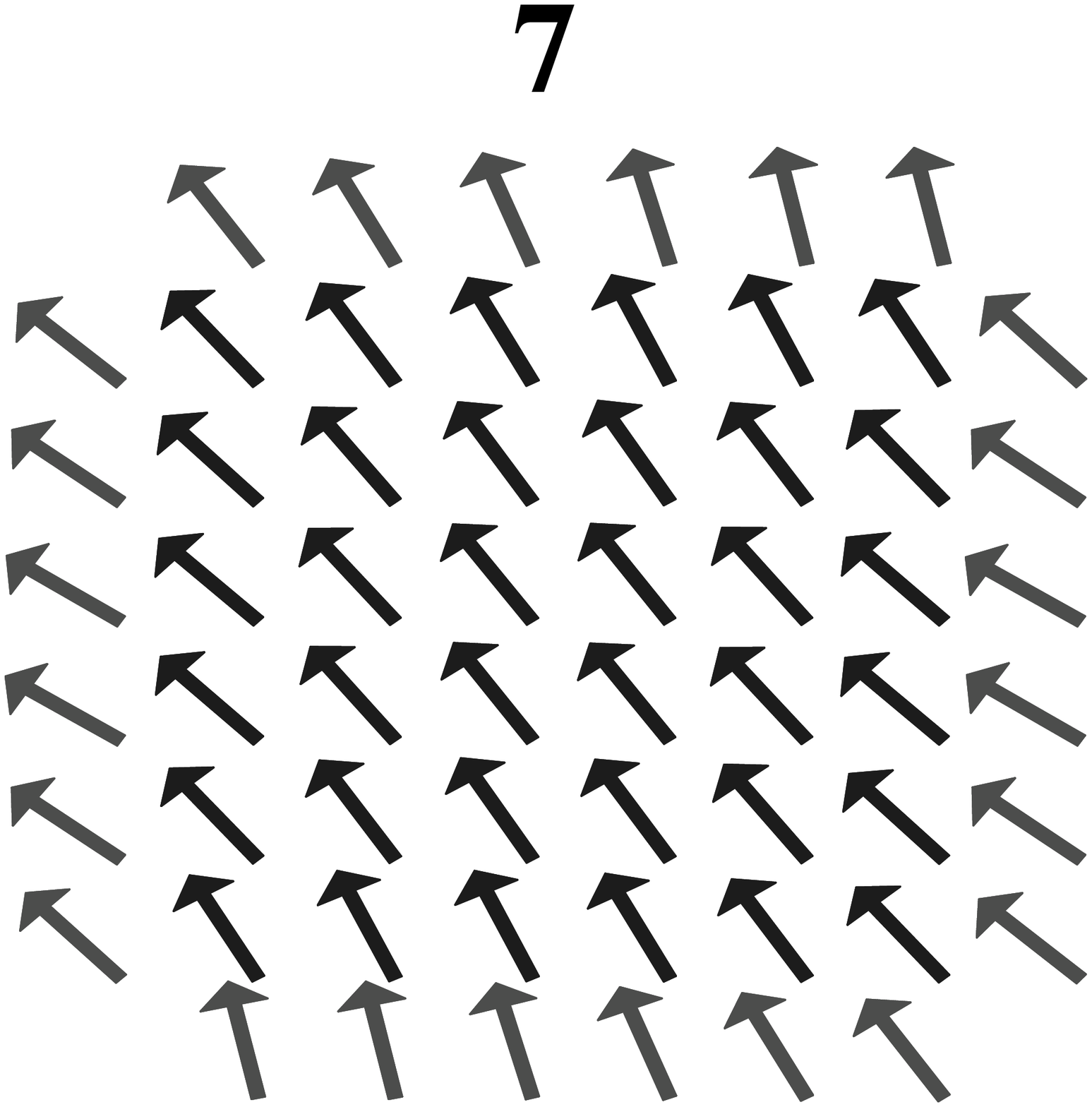}%
\hspace{0.25cm}
\includegraphics[angle=-90,width=2.5cm]{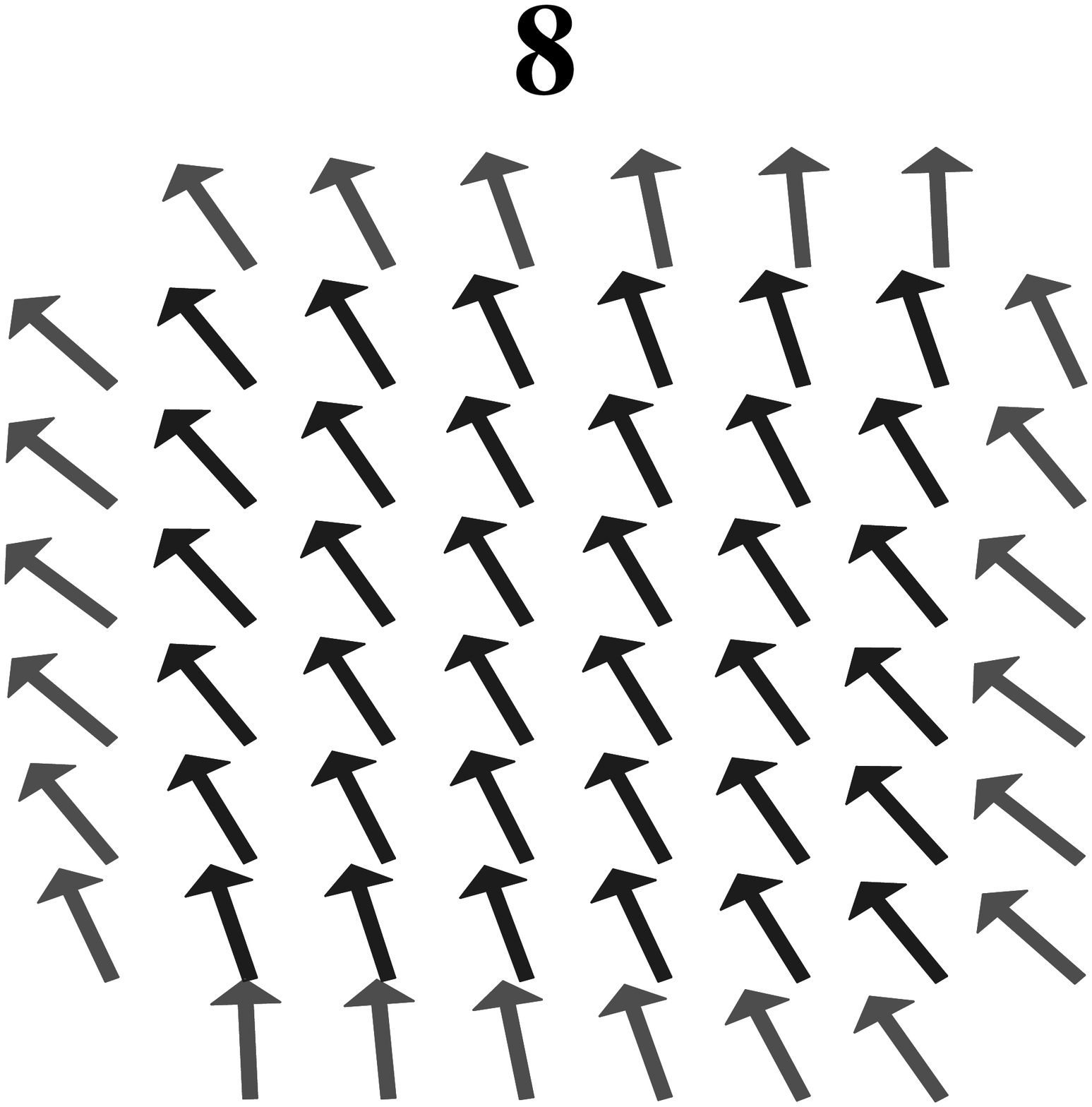}\\[0pt]
\vspace{0.5cm}
\includegraphics[angle=-90,width=2.5cm]{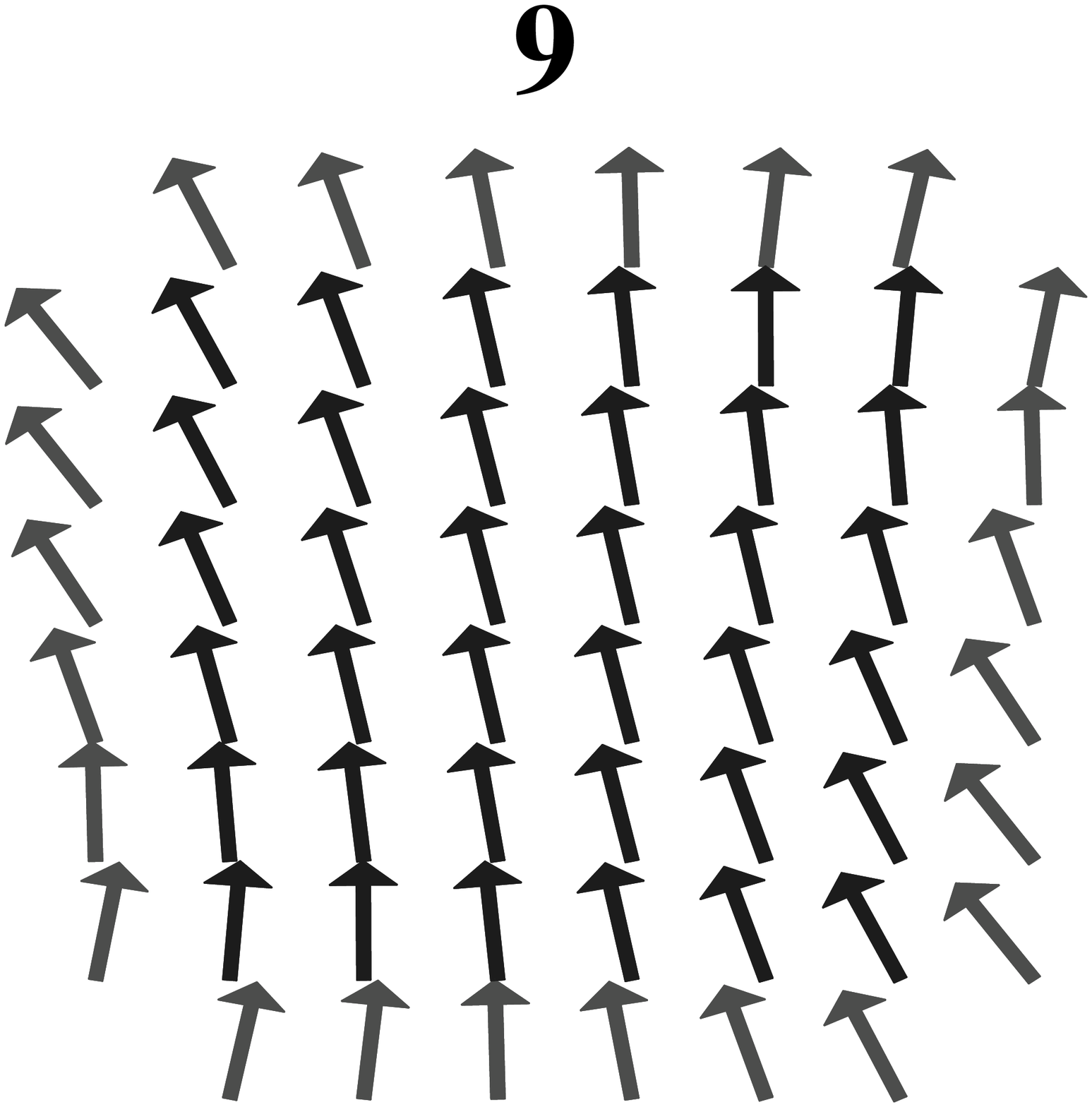}\hspace{%
0.25cm} \includegraphics[angle=-90,width=2.5cm]{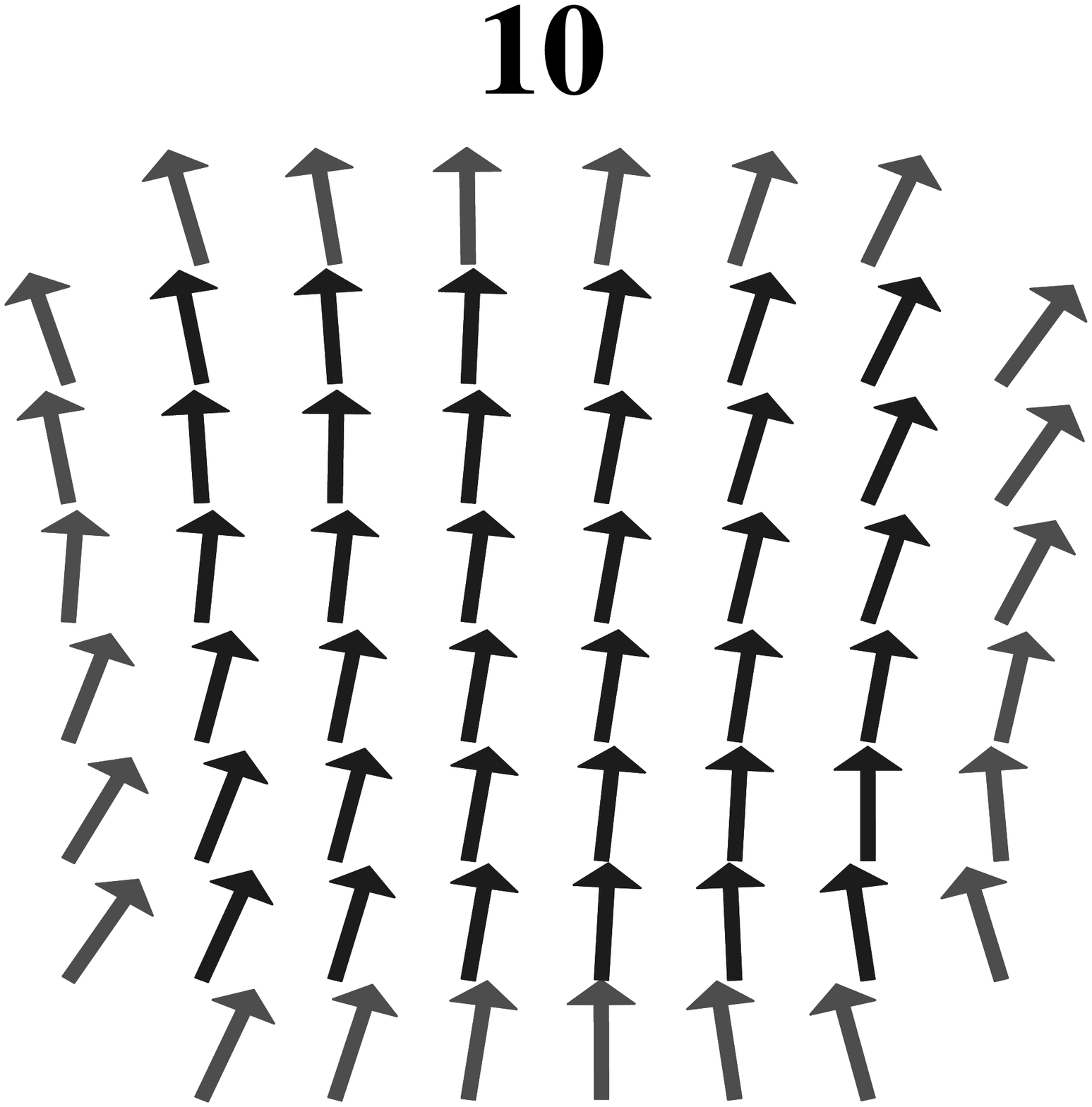}%
\hspace{0.25cm}
\includegraphics[angle=-90,width=2.5cm]{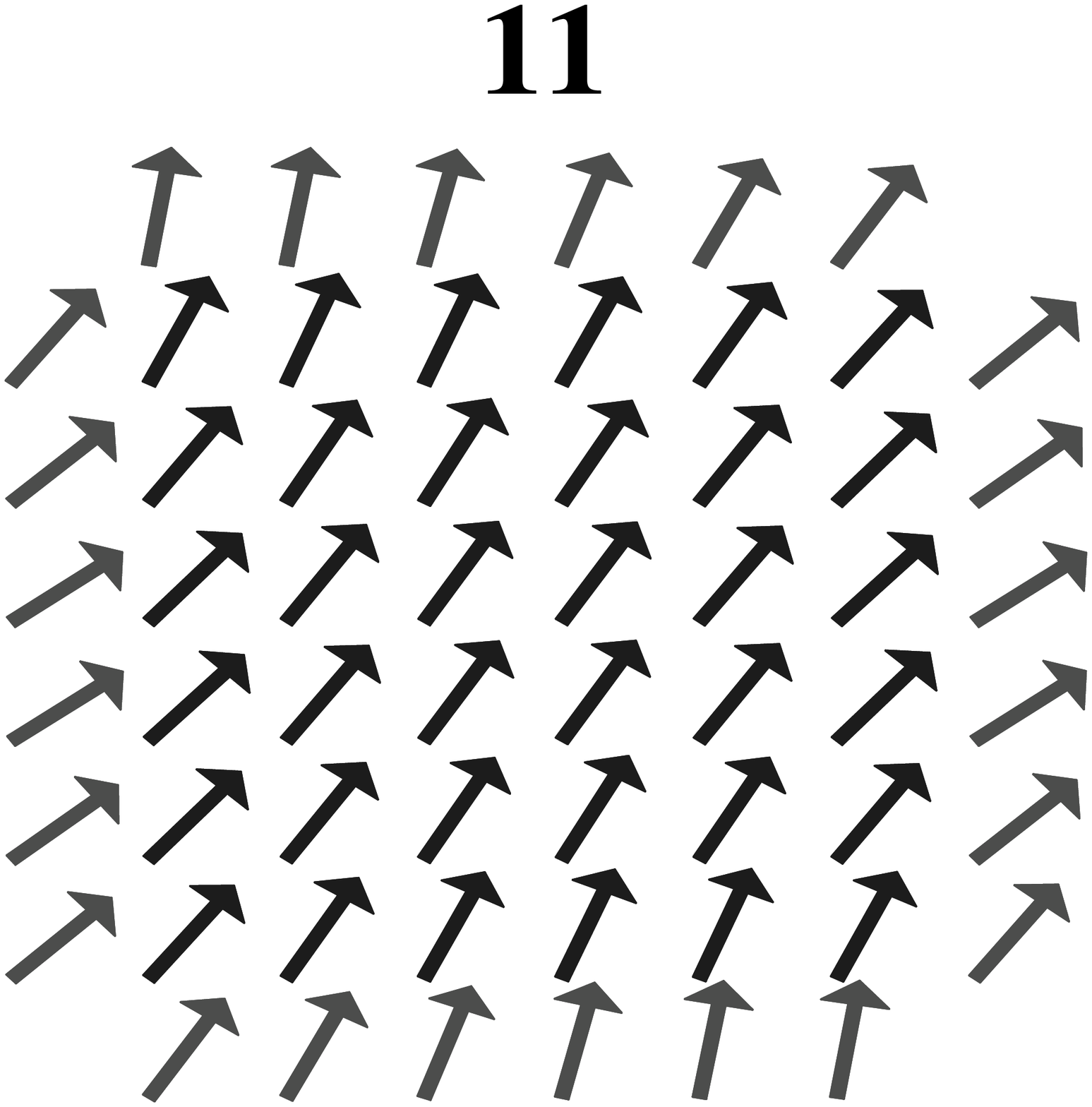}\hspace{%
0.25cm} \includegraphics[angle=-90,width=2.5cm]{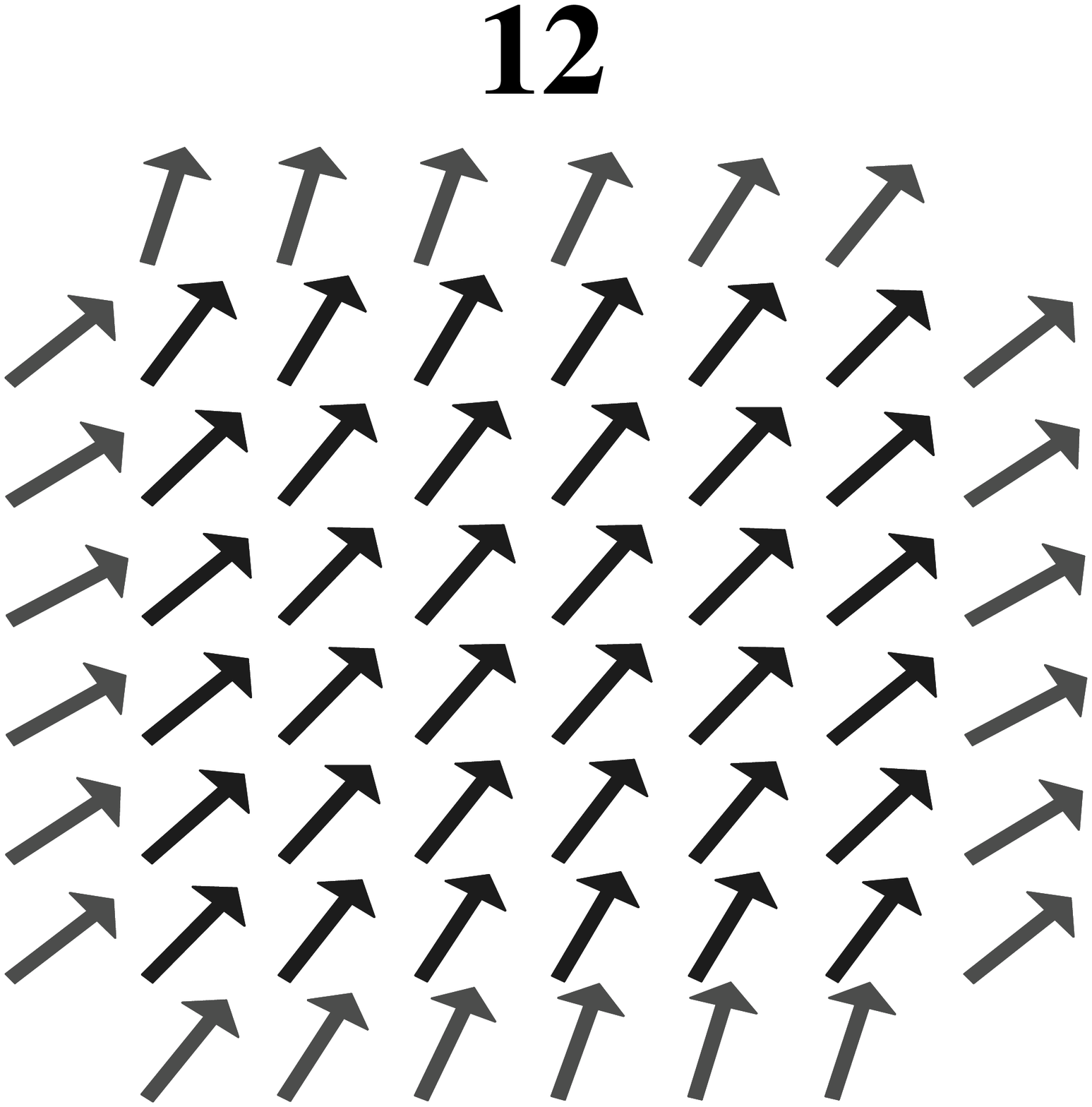}
\end{center}
\caption{Magnetic structures corresponding to the middle plane of the
particle. The parameters are the same as in Fig.~\ref{loops_full_vs_minor}.
The indices on top of the structures correspond to field values shown in
Fig.~\ref{loops_full_vs_minor}.}
\label{ks30n11a45_ih0}
\end{figure*}
%
For later discussion, in Fig.~\ref{ks30n11a45_ih0} we present the magnetic
structures corresponding to the middle plane of the particle where the
indices on top of the structures correspond to those shown in Fig.~\ref
{loops_full_vs_minor}. 
%
\begin{figure}[floatfix]
\begin{center}
\includegraphics[width=8cm]{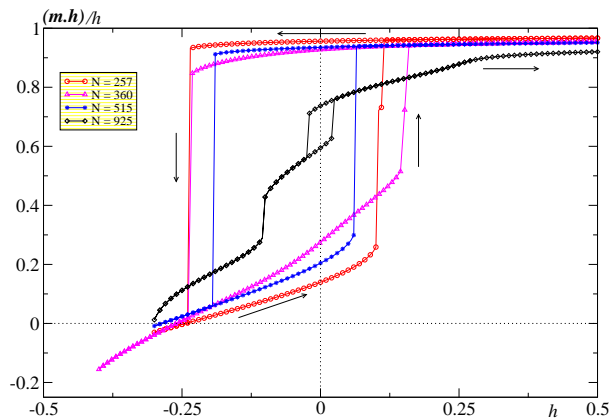}
\end{center}
\caption{(Color online) Minor hysteresis loops for the particle diameters $D=9,10, 11$ and $13$ in the units of the atomic spacing, corresponding to the total number of atoms indicated in the legend. The other parameters are as in Fig.~\ref{loops_full_vs_minor}.}
\label{loops_minor_size}
\end{figure}
%
In Fig.~\ref{loops_minor_size} we see that the bigger is the particle, and
thereby the smaller is the surface contribution, and the narrower is the
minor loop, which implies that the non-collinearities caused by surface
anisotropy are indeed responsible for the minor loop.
%
\begin{figure*}[floatfix]
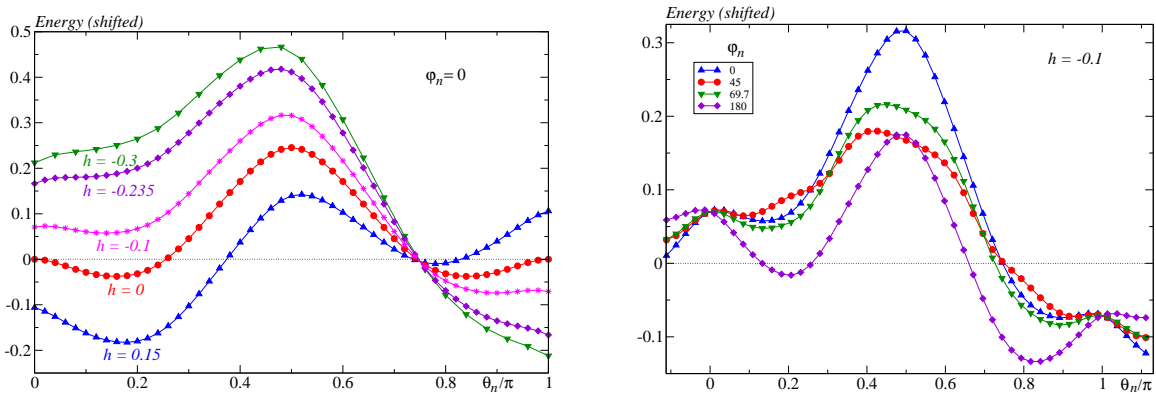

\begin{center}
\includegraphics[width=7.3cm]{enscape_field.eps}\hspace{1cm}
\includegraphics[width=6.9cm]{enscape_phi.eps}
\end{center}
\caption{(Color online) Left: Energy as a function of the polar angle $\theta_{n}$ for $%
\phi_n=0$ and different field values. Right: Energy as a function of the
polar angle $\theta_n$ for $h=-0.1$ and different values of the azimuthal
angle $\phi_n$. }
\label{enscape_field_phi}
\end{figure*}
%
The left graph in Fig.~\ref{enscape_field_phi} shows how the initial energy
minimum corresponding to the initial state evolves when the field is varied
along the descending branch of the hysteresis loop in Fig.~\ref
{loops_full_vs_minor}. The graph on the right shows the change in the
energyscape when the azimuthal angle is varied at the field $h=-0.1$.
Likewise, in Fig.~\ref{mag_cc} we plot the evolution in field of the
Cartesian coordinates of the net magnetization.
%
%
Upon examining Figs.~\ref{enscape_field_phi}, \ref{mag_cc}, we see that the
orientation of the particle's net magnetization varies according to the
following scenario: it starts in the minimum defined by the positive
saturation state, i.e., $(\theta_n\sim 35^{\circ },\varphi_n=0)$. As the
field decreases towards negative values, this orientation drifts towards the
$z$ axis ($\theta_n\sim 14^{\circ },\varphi_n=0$), as confirmed by the
increasing $m_{z}$ in Fig.~\ref{mag_cc} (circles). This continues until the
field reaches the value $\sim -0.235$. Note that the net magnetization stays
in the $x-z$ plane containing the field direction. Then, the net
magnetization jumps to the direction $(\theta_n\sim 38^{\circ
},\varphi_n=\pi )$. This occurs through a saddle point in the $\varphi_{n}$
direction as can be seen in Fig.~\ref{enscape_field_phi} (right). As the
field further decreases, this direction goes towards the $x$ axis, but still
with $\varphi_{n}=\pi $. When we cycle back from the field value $h=-0.4$
across the value $-0.235$, the direction of $\mathbf{m}$ goes through
another saddle point and so does not jump back to $(\theta_n\sim 14^{\circ
},\varphi_n=0)$ but continues towards the $z$ axis until the field reaches
the value $h=0.15$ at which it jumps to the state marked by symbols within
circles in Fig.~\ref{mag_cc}.
In fact, this is the only state in which the net
magnetization goes out of the $x-z$ plane. From the latter state, the
direction of the net magnetization jumps back to the initial state.

It is clear that this scenario should be dependent on the energyscape, and
thereby on the particle's size (see Fig.~\ref{loops_minor_size}) and surface anisotropy (constant and model).
Nonetheless, it shows that surface anisotropy induces more minima and saddle points in the
energyscape, and thereby more paths along which the direction of the net
magnetization can evolve as the field is varied.

\section{\label{EOSP}Effective OSP}

\begin{figure*}[floatfix]
\includegraphics[width=8.5cm]{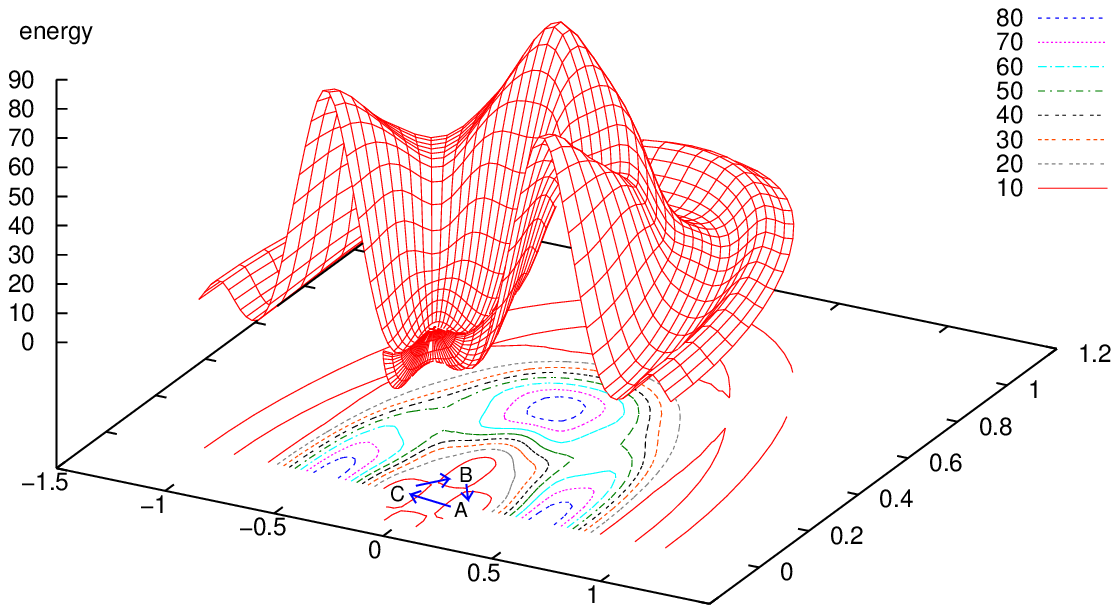}
\includegraphics[width=6.5cm]{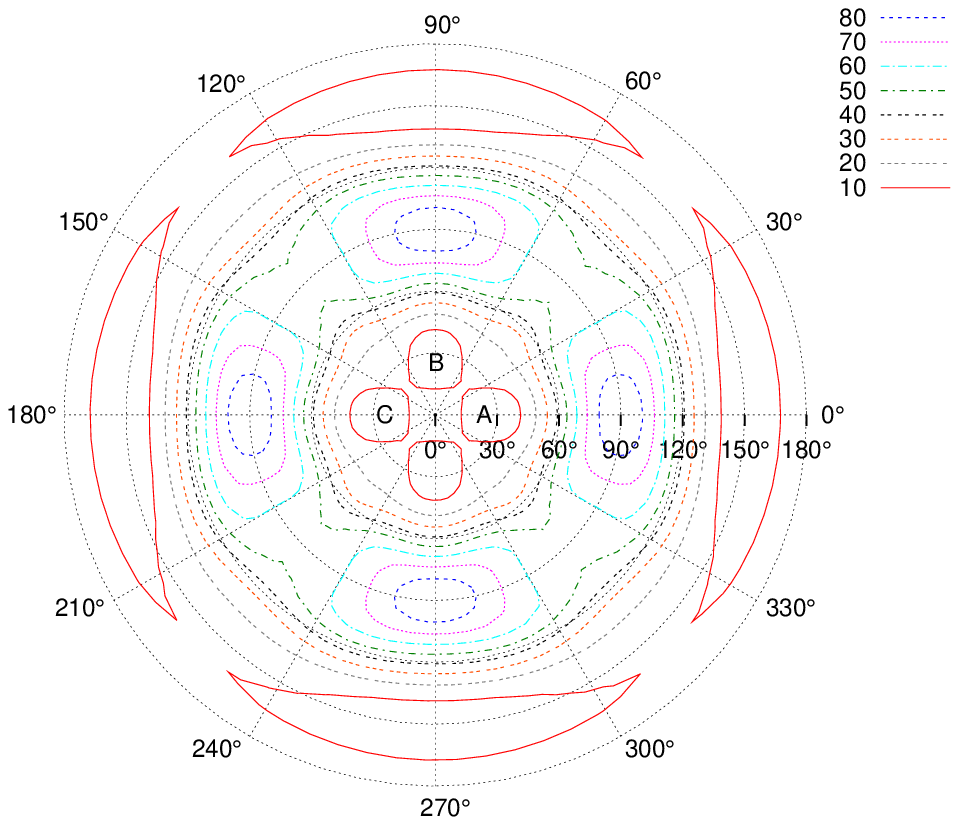}
\caption{(Color online) Left: Plot of $(\left(\theta_n/\pi\right)\cos \varphi_n ,\left(\theta_n/\pi\right)\sin\varphi_n ,E)$ for the same parameters as before. Right: Contour plot of
the energy in the plane of polar coordinates $\rho =\theta_n$ and $%
\theta_n^{\prime }=\varphi_n$.}
\label{energy_contour_plot}
\end{figure*}

The results discussed above suggest that the MSP may be represented by an
effective magnetic moment [as is done in the so called macro-spin
approximation \cite{wernsdorfer01acp, bonetetal99prl}] which is just the net magnetic moment
of the particle in an effective anisotropy potential. Indeed, in \cite
{garkac03prl} it was shown that when the surface anisotropy is smaller than
exchange, in the absence of core anisotropy, the surface anisotropy
contribution to the particle's energy is of $4^{\mathrm{th}}$ order in the
net magnetization components and second order in the surface anisotropy
constant, and is proportional to a surface integral. Now, we may ask the
question as to what happens in the presence of core anisotropy. The answer
is obtained upon computing the $2D$ and $3D$ energyscape for the MSP as
explained earlier [see Eq. (\ref{FFuncDef}) et seq]. The results are shown
in Fig.~\ref{energy_contour_plot}.
%
%
They represent the $2D$ and $3D$ energyscape of the particle at zero field
as a function of the direction $\mathbf{m}$ of the net magnetization. The
locus of $\mathbf{m}$ (the unit sphere) has been mapped onto a plane using
the azimuthal equidistant projection: the spherical coordinates $\theta_n$
and $\varphi_n$ become the polar coordinates $\rho =\theta_n$ and $%
\theta_n^{\prime}=\varphi_n$ of the plane. This is the projection used for
the UN logo. It preserves the aspect of the ``north pole'' region, i.e.\ the
region around $\theta_n=0$, which is an easy direction for the core
anisotropy. The $2D$ graph on the right shows that the high maxima seen in
the $3D$ graph are indeed at $\theta_n = \pi/2$ and that the ``beans'' on the
border are the antipodes of the minima around $\theta_n = 0$ and at $A, B, C$. As can be seen from these plots, the effect of the non-collinearities of the magnetization on
the energy is to split the minimum at $\theta_n=0$ into four minima at $%
\theta_n\sim 28^{\circ }$ and $\varphi_n=0,\pm\pi/2, \pi$. These minima
are connected by saddle points at $\varphi_n=\pm \pi /4$ and $\pm 3\pi /4$
and the point at $\theta_n=0$ becomes a small local maximum. The four minima
exist over a finite range of applied field, although their positions change
continuously as a function of the field. They can explain the minor loop of
Fig.~\ref{loops_full_vs_minor} as follows: In the upper branch of the loop
(structures $1, 2$ and $3$ in Fig.~\ref{ks30n11a45_ih0}), the magnetization
is in the minimum labeled $A$. At $h\sim -0.235$, it jumps to minimum $C$
(structure $4$). It stays in this minimum (structures $4-9$) until the
field is swept up past $h\sim 0.15$, at which point it jumps to the minimum $%
B$ (structure $10$) and then back to minimum $A$ (structures $11$ and $12$).
The only minimum with the magnetization out of the $x-z$ plane, namely $B$
[marked by circles in Fig.~\ref{mag_cc}], is reached when the field is swept
up, but not when it is swept down. This asymmetry is due to the fact that
the field is applied in the direction $\theta_n=\pi /4,\varphi_n=0$, which
is closer to the minimum $A$ than to the other three.

This suggests that the energy of such an MSP can be modeled by that of a
one-spin particle with the net magnetization $\mathbf{m}$ in a potential
containing a uniaxial and cubic anisotropy terms, i.e., up to a constant,
\begin{equation}
\mathcal{E}_{\mathrm{eff.}} = -K_{\mathrm{uni}
}\,m_{z}^{2}+K_{\mathrm{cub}}(m_{x}^{4}+m_{y}^{4}+m_{z}^{4}).
\label{UniaxialCubicEnergy}
\end{equation}

\begin{figure*}[floatfix]
\begin{center}
\includegraphics[width=4.5cm, angle=-90]{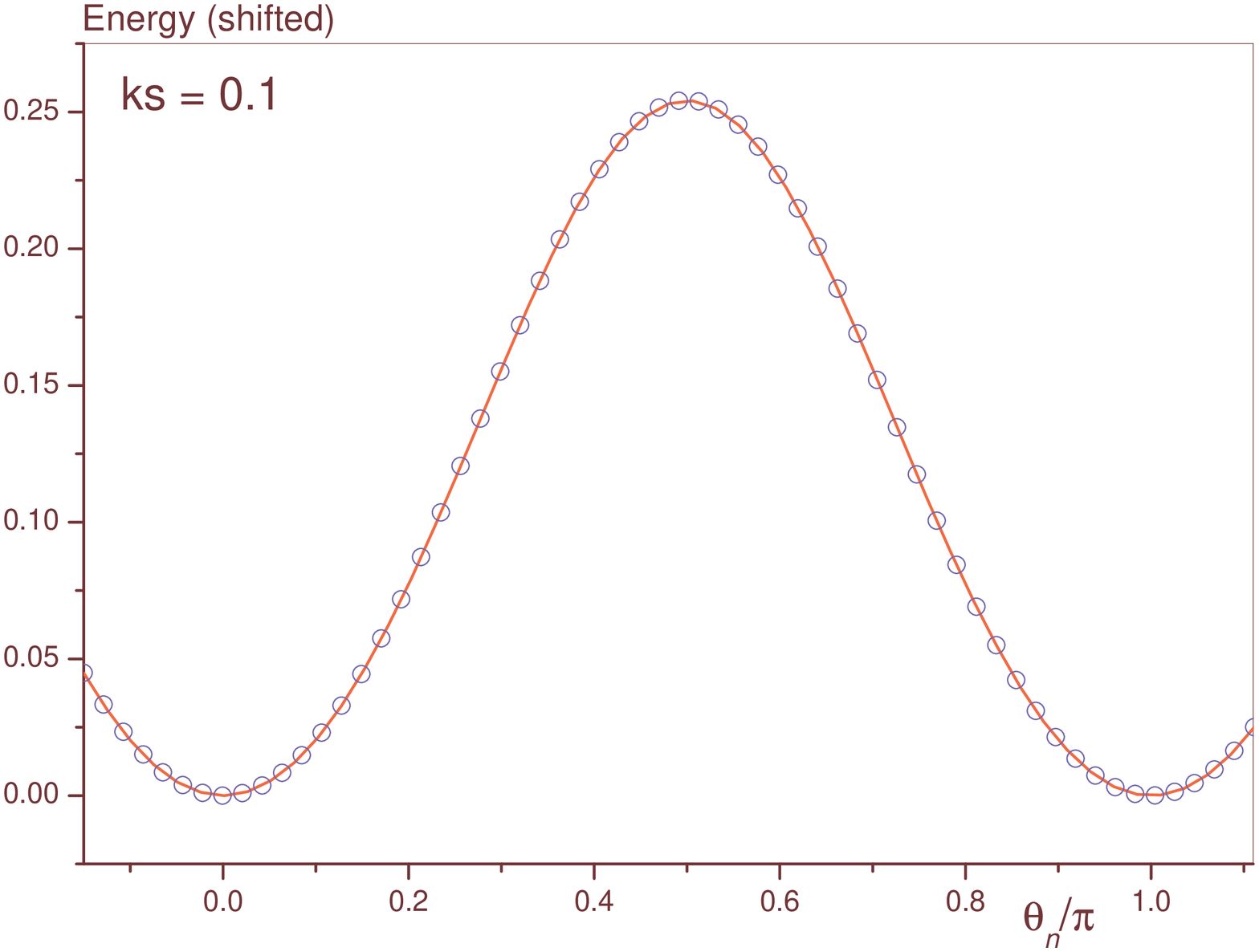}
\includegraphics[width=4.5cm, angle=-90]{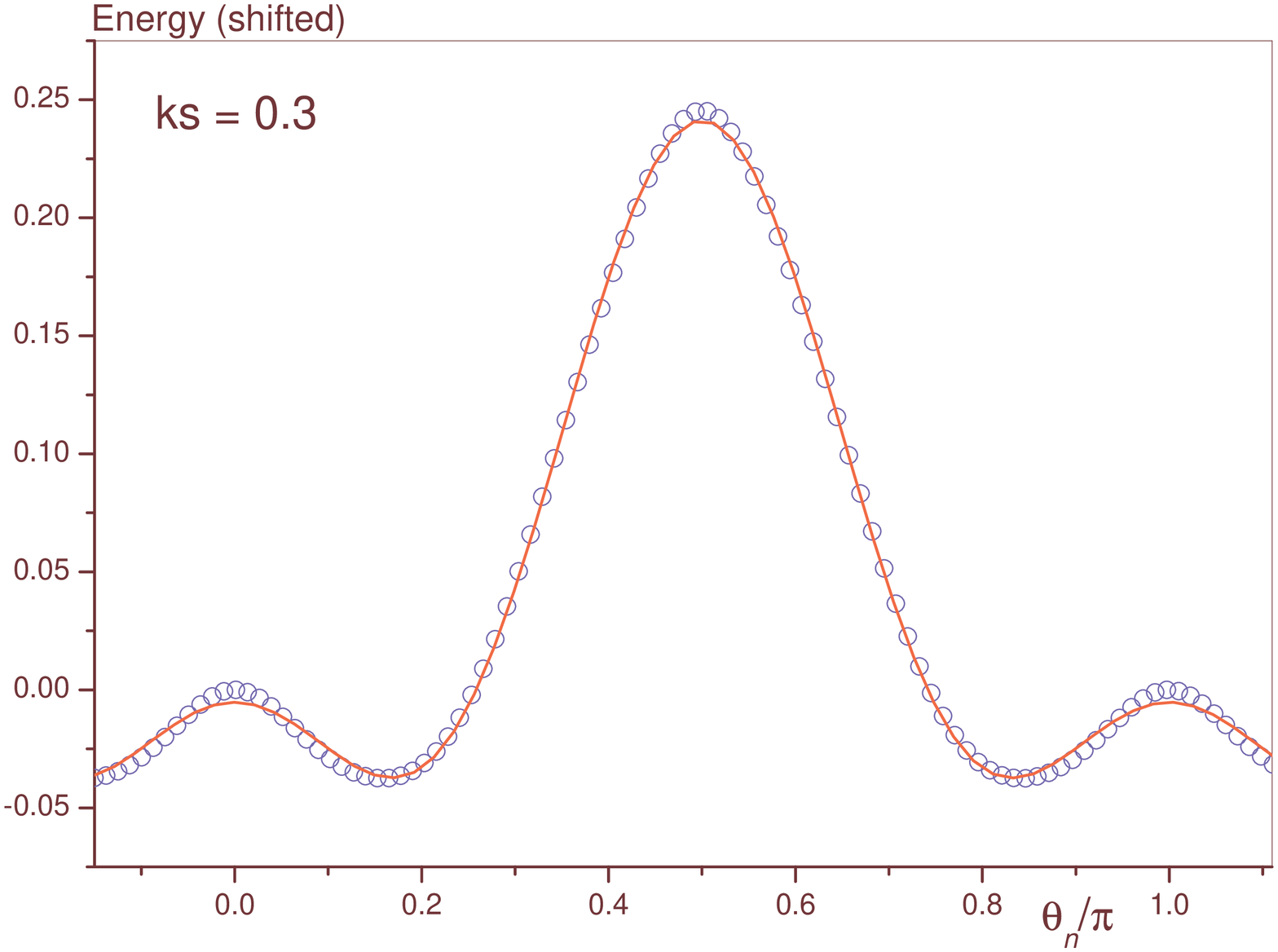}
\includegraphics[width=4.5cm, angle=-90]{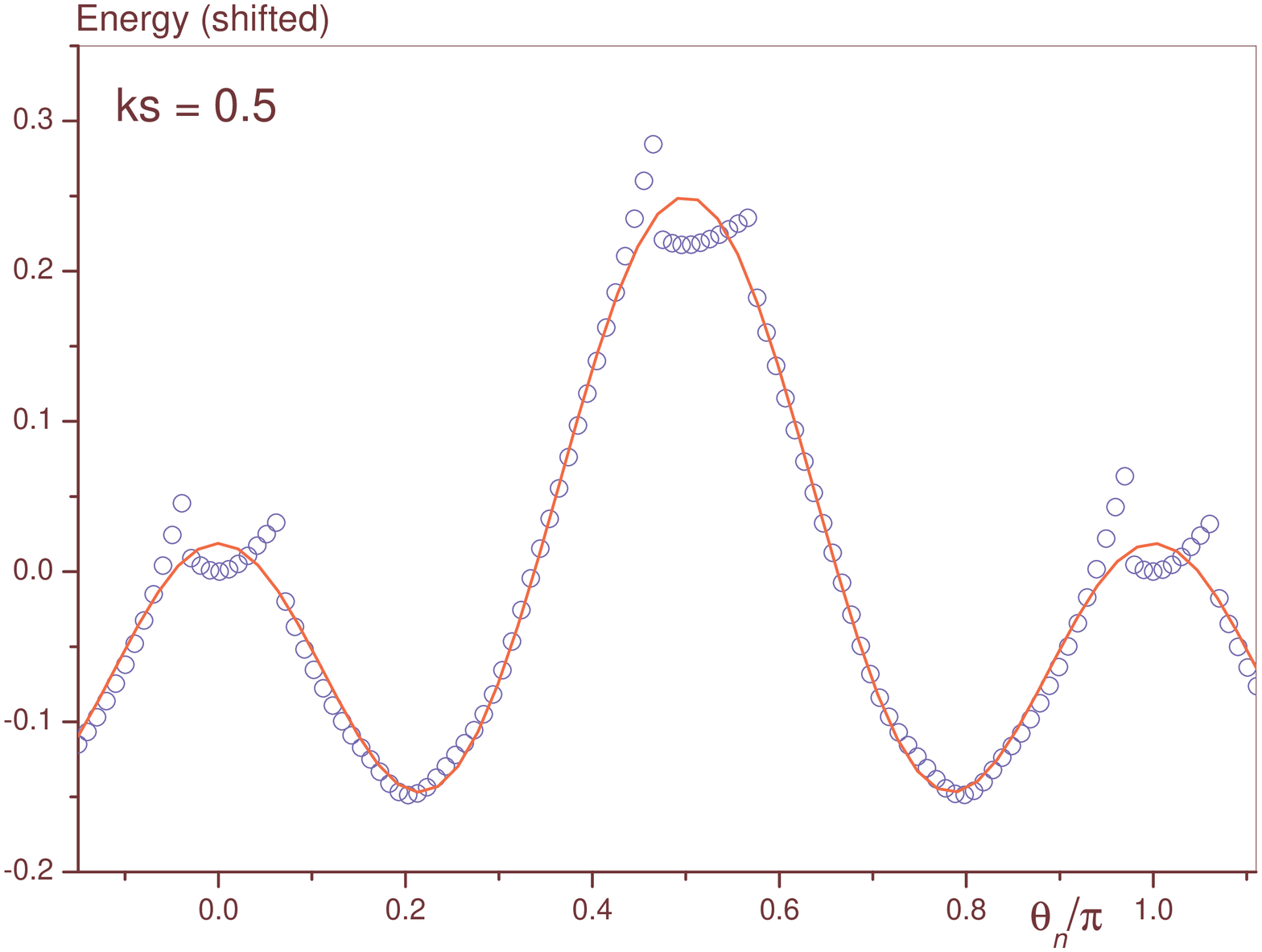}
\end{center}
\caption{(Color online) $2D$ energyscape for the spherical particle of Fig.~\ref{loops_full_vs_minor} with varying surface anisotropy constant $k_s$. The symbols are results of the numerical calculations for the MSP, and the full lines are fits using Eq. (\ref{UniaxialCubicEnergy}).
$\phi_n=0$ and $h=0$.}
\label{enscape2D_h0_varyks}
\end{figure*}

Indeed, Fig.~\ref{enscape2D_h0_varyks} (first panel, $k_s=0.1$) shows that for very small $k_s$ the energyscape of an MSP is well recovered by the effective energy in Eq.~(\ref{UniaxialCubicEnergy}).
%
%
As $k_s$ increases [see middle panel, $k_s=0.3$], some deviations start to be seen, and for relatively larger values of $k_s$ a fit with Eq.~(\ref{UniaxialCubicEnergy}) is no longer possible. In fact, in this regime strong deviations from collinearity develop, especially near maxima and saddle points, as can be seen on the last panel ($k_s=0.5$) in Fig.~\ref{enscape2D_h0_varyks}. In fact, as mentioned earlier, in this case the Lagrange-parameter method in Eqs.~(\ref{FFuncDef}, \ref{LLEqs}) fails because the magnetic state of an MSP can no longer be represented by a net magnetization.
Next, fitting the $3D$ energy plotted in Fig.~\ref{energy_contour_plot} with expression (\ref{UniaxialCubicEnergy}) leads to $K_{\mathrm{uni}}\simeq 0.0049$ and $K_{\mathrm{cub}}\simeq 0.0036$. The uniaxial-anisotropy term stems mainly
from the core contribution, though with a small contribution from the surface. Indeed, $K_{\mathrm{uni}}=(N_{c}/\mathcal{N} )\times K_{c}$, where $N_{c}$ is the number of core spins ($=184$ here), leading to $K_{c}\simeq 0.0096$, which is close to the constant initially used. This simple fitting shows that even for the seemingly strong surface anisotropy ($k_{s}=0.3$) and large surface-to-volume ratio ($\sim 0.49$ here), the surface contribution to the effective energy remains smaller than the contribution of the uniaxial core anisotropy ($K_{\mathrm{cub}} \lesssim K_{\mathrm{uni}}$).

\begin{figure}[floatfix]
\includegraphics[width=9cm]{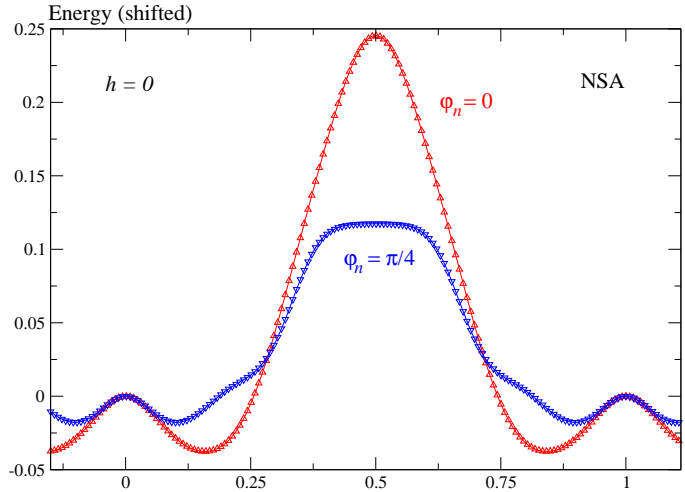}
\caption{(Color online) Energy as a function of the polar angle $\theta_{n}$ for a spherical particle with uniaxial anisotropy in the core and N\'eel anisotropy on the surface. $\phi_n=0,\pi/4$ and $h=0$. The other parameters are as in Fig.~\ref{loops_full_vs_minor}.}
\label{NSA360kc001ks03_p0p45}
\end{figure}

The results presented hitherto are all for a spherical particle with uniaxial anisotropy in the core and TSA on the surface.
Fig.~\ref{NSA360kc001ks03_p0p45} is a plot of the $2D$ energyscape for a spherical particle with uniaxial anisotropy in the core, as before, but now with NSA on the surface.
%
%
It is clear that the cubic-anisotropy features are also seen in the case of NSA, namely that i) the energy minima are not along the directions ($\theta_{n}=0,\pi$) of the core easy axis, these directions having become local maxima, as discussed in the case of TSA, and ii) there is a clear dependence on the azimuthal angle $\phi_n$.

The results above agree and complement those of Ref.~\onlinecite{kacdim02prb} and \onlinecite{kacmah04jmmm}, where it was shown that for the TSA and NSA there exists a (different) critical value of the surface anisotropy constant that separates i) the OSP Stoner-Wohlfarth regime of coherent switching and ii) the MSP regime where the strong spin non-collinearities invalidate the coherent mechanism, and the particle can no longer be modeled by an effective OSP.
Obviously, for very small surface anisotropy the cubic contribution becomes negligible [see the first panel in Fig.~\ref{enscape2D_h0_varyks} for $k_s=0.1$] and the Stoner-Wohlfarth OSP model provides a good approximation to the many-spin particle.
Accordingly, some experimental macroscopic estimations of the surface anisotropy constant yield, e.g., for cobalt $K_s/J\simeq 0.1$ \cite{skocoe99iop}, for iron $K_s/J\simeq 0.06$ \cite{urquhartetal88jap}, and for maghemite particles $K_s/J\simeq 0.04$ \cite{perrai05springer}.
However, one should not forget that this effective constant depends on the particle's size, among other parameters such as the material composition, and for, e.g. a diameter of $2$ nm we may expect higher anisotropies.

The present results have been obtained for a field directed at an angle of $\pi/4$ with respect to the core easy axis. In Ref.~\cite{kacdim02prb} we showed that the effect of varying this angle from, e.g. $0$ to $\pi/4$, was to decrease the abovementioned critical value of the surface anisotropy constant.
For simplicity, in this work we only consider this typical value as it is half-way between the singular values of $0$ and $\pi/2$, which is also useful for particle  assemblies with randomly-distributed anisotropy easy axes.

Finally, we would like to mention that more extensive calculations \cite{kachkachietal06prep} have shown that similar features are also observed for other crystal structures (fcc, bcc, etc.).

\section{\label{conclusion}Conclusion}
We have shown that the behavior of a many-spin particle with core and (weak) surface anisotropy can be modeled by that of an effective macro-spin particle whose net magnetization experiences an effective potential energy containing a $2^\mathrm{sd}$- and $4^\mathrm{th}$-order contributions.
However, a question remains open to further investigations: how to distinguish this surface-induced $4^\mathrm{th}$-order contribution from the (usually weak)
cubic anisotropy found in many materials in addition to the $2^\mathrm{sd}$-order contribution?
In Ref.~\onlinecite{jametetal01prl} a fit to the $3D$ Stoner-Wohlfarth astroid leads to a $4^\mathrm{th}$-order term an order of magnitude smaller than the $2^\mathrm{th}$-order one.

The results of the present work could be useful for investigating the thermally activated reversal of a small nanoparticle. Indeed, the effective OSP with the energy (\ref{UniaxialCubicEnergy}) allows us to include surface effects, though in a phenomenological manner, while offering a considerable simplification as compared to the initial many-spin system.
However, one should note that the effective potential contains a quartic term in the net magnetization components, which renders the analysis of the energyscape somewhat more involved but still feasible \cite{kalmykov00prb}, as opposed to the case of a many-spin particle.
In fact, for small enough surface anisotropy for which the effective energy (\ref{UniaxialCubicEnergy}) holds best, the minima are mainly defined by the uniaxial anisotropy and the effect of the cubic-anisotropy contribution is then to modify the loci of the saddle points and their number. This obviously renders the calculation of the relaxation rate rather involved \cite{kalmykov00prb}.

In the general case of arbitrary surface anisotropy intensity, one has to tackle the problem of computing the relaxation rates for a many-spin system with all the ensuing complexities. Such difficulties are already encountered, in this context, in the much simpler situation of two exchange-coupled spins \cite{kac03epl, titovetal05prb}.
Obviously, the general situation can only be dealt with through numerical approaches.
%

\end{document}